# Extension of the Hertz model for accounting to surface tension in nano-indentation tests of soft materials

## C. Fond* — O. Noël** — M. Brogly***

*Institut Charles Sadron (CNRS)
6 rue Boussingault, F67083 Strasbourg
cfond@ics.u-strasbg.fr

**Université du Maine
Avenue Olivier Messiaen, F72085 Le Mans
olivier.noel@univ-lemans.fr

***Institut de Chimie des Surfaces et Interfaces (CNRS)
15, rue Jean Starcky, BP 2488, 68057 Mulhouse cedex
m.brogly@uha.fr

ABSTRACT: The contact between a spherical indenter and a solid is considered. A numerical finite element model (F. E. M) to taking into account the surface tension of the solid is presented and assessed. It is shown that for nano-indentation of soft materials, the surface tension of the solid influences significantly the reaction force due to indentation. The validity of the classical Hertz model is defined. In very good approximation, the force vs. indentation depth curve can be fitted by a power law function $F = a\ \delta^b$ where $F$ denotes the force acting on the indentor, $\delta$ the indentation depth, $a$ and $b \in ]1, 1.5]$ are constants depending on the materials and the size of the indentor.

RÉSUMÉ: Le contact entre un indenteur sphérique et un solide élastique est considéré. Un modèle par éléments finis permettant de prendre en compte la tension de surface du solide est présenté et validé. Il est montré que pour la nano-indentation de matières souples, la tension de surface du solide peut influencer considérablement la courbe force vs. profondeur d'indentation. Il est ainsi définit quand le modèle de Hertz classique ne convient plus. En très bonne approximation, la courbe force vs. profondeur d'indentation peux être décrite par une croissance allométrique $F = a\ \delta^b$ où $F$ désigne la force agissant sur l'indenteur, $\delta$ la profondeur d'indentation, $a$ et $b \in ]1, 1.5]$ sont des constantes qui dépendent du matériau et du rayon de pointe.

MOTS-CLÉS: tension de surface - nano-indentation - contact - surface - matière molle - éléments finis - dureté

## 1. Introduction

Indentation tests are powerful techniques to measure, at least, the elastic moduli of materials.[1,2] However, considering soft materials, such measurements using nano-indentor generally give an overestimate of the Young's modulus measured at larger scales in uniaxial tension[3,4,5]. In order to distinguish mechanical properties at the surface and in the bulk of a given material, it is convenient to evaluate the effect of surface forces on the mechanical response. To achieve this objective, this paper quantifies the effect of the surface tension of soft material on the reaction force due to indentation for nano-indentation tests through finite elements simulations. Indeed, it is not easy to derive an analytical solution for the elastic problem coupled with surface tension while, the finite elements method gives numerical solutions for such problems[6]. Moreover finite element simulation has the advantage to allow to consider various mechanical behaviours and shapes of the indentor. However, this paper only concerns spherical indenters and elastic materials.

In preliminary to the simulation, one must consider that the surface tension, t, varies with the surface curvature only at very low radius, typically less than one nanometre[7]. For our purpose, t should remain constant. It is well known that the energy associated to the surface tension can be equivalently seen as an energy corresponding to a change in area or as the work done by a tension[8], usually called Laplace tension.

## 2. Mechanical model

### 2.1. Finite element model

A simple way is proposed herein to account for surface tension using a standard finite element software, Cast3M[©9]. A stretched "membrane" linked to the surface of the solid is considered. The Figure 1 shows schematically the axisymmetrical (the $x_1$ axis is equivalent to $x_2$ axis) mesh of the solid, $x_3$ being the axis of symmetry. It is obvious that such computations must be performed with large displacements analysis[10] even at small strains so that the tension in the stretched "membrane" can induce a reaction force along the $x_3$ axis. The spherical indentor is assumed to be perfectly rigid. Therefore, the indentor corresponds to a region where the material can not penetrate. A step of computation corresponds to an increase of indentation depth. The boundary conditions are defined as following: imposed displacements for nodes in contact with the indentor and free displacements outside the contact region for the surface nodes. These conditions are actualised at each step of the computation.

The axisymmetric finite elements model (FEM) of the membrane located at the surface uses four nodes isoparametric elements. This membrane must be sufficiently thin to have a negligible effect on the mechanical response in absence of surface tension, i. e. $t = 0$. As far as possible, the tension in the membrane must be constant to simulate a constant surface tension. A constant bi-axial constraint $\sigma_0$ in the $x_1$ and $x_2$ axis are then imposed in these elements. Their mechanical and geometric properties are set to $E_t = E$, $v_t = 0$, $e < r / 1000$, $\sigma_0 > 10^4$ E where $E_t$ is the Young's modulus, $v_t$ the Poisson's ratio, e the thickness of the membrane and $\sigma_0$ the pre-constraint in the radial and ortho-radial directions. The surface tension value in the undeformed state, i. e. a planar membrane, is denoted $t_0$ and equal to $e.\sigma_0$. In order to avoid numerical convergence problems, the nodes initially aligned along the $x_3$ axis of each element constituting the "membrane" remains aligned along the $x_3$ axis for each step of the computation, as shown in Figure 2 (left).

The boundary conditions at $x_1 = D$ and $x_3 = -h$ impose that displacements are null, i. e. $u_1(x_1 = D) = u_3(x_3 = -h) = 0$. Practically, boundaries are set sufficiently far from the indenter to avoid any border effect. This means that D / r and h / r are sufficiently large so that it makes no significant difference to increase D or h. Moreover, it makes also no significant difference to impose $u_1 = 0$ and $u_3 \neq 0$ (the corresponding nodal force $f_3$ is thus zero) along the $x_3$ axis and $u_1 \neq 0$ ($f_1 = 0$) and $u_3 = 0$ along the $x_1$ axis.

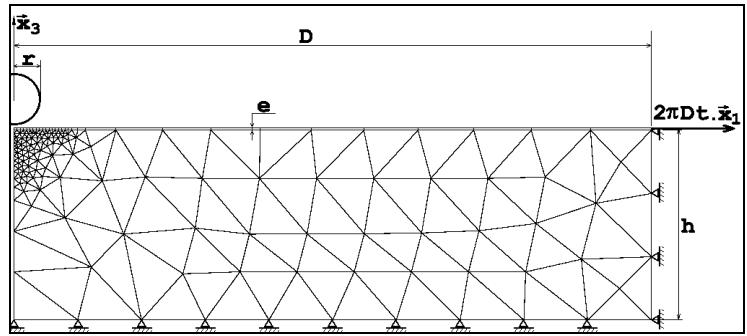

**Figure 1.** *Schematic finite elements axisymmetrical ($x_1 = x_2$) mesh showing the solid substrate, the spherical indenter and the stretched membrane associated to the surface, acting like surface tension.*

## 2.2. Assessment from numerical analysis

Since the pre-constraint $\sigma_{11} = \sigma_{22} = t_0 / e$ is constant, using a standard procedure of the finite element software, a rotation of the element induces an increase of the tension along the element. Figure 2 (right) shows the evolution of the tension in the surface element versus the angle of rotation with the r axis. Since practically this angle is low for all computations presented below, always less than 15°, the surface tension or equivalently the constraint in the element remains almost constant[6]. Indeed, for moderate indentation depths, i. e. $\delta / r < 1$, the tension varies less than 3 %.

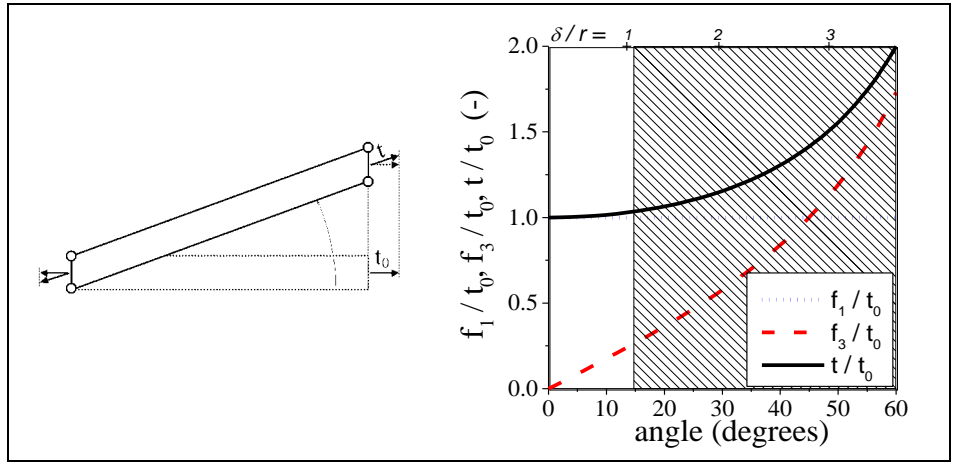

***Figure 2.*** *(left) Kinematic and equilibrium of the four nodes elements used to simulate the surface tension (scheme not at scale). (right) Evolution of the surface tension t with the rotation of the surface element.*

The evolution of the surface tension inside an element is representative of the local discrepancy and informs about the maximum error on the surface tension located at the limit of the contact area where the rotation of the element is maximum. Nevertheless, it is convenient to quantify the mean error by considering the work done by the surface tension, i. e. the retrieved surface energy, considering the inflation of a soap bubble, i. e. a spherical stretched membrane, representative of a spherical indentor. The Figure 3 shows the work done by the surface tension, normalized by the surface energy, below a spherical indentor versus curvature corresponding to a contact area comprise between 0 and r. This test assesses the ability of this simple numerical model to quite accurately account for a constant surface tension for contact area where a / r is typically less than 0.4.

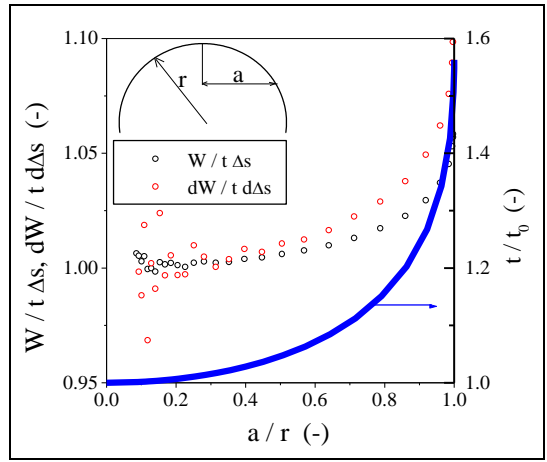

**Figure 3.** *Evolution of surface energy and the maximal surface tension in soap bubble vs. curvature.*

Still considering only the surface tension, like in the soap bubble problem, numerical results from the FEM are compared in Figure 4 to the analytical solution[6] from eq. A1. 2 derived in Appendix 1. Results are in good agreements since for $\delta / r < 0.1$ the error is less than 3 %. It is noticeable that the fit of the FEM curve with allometric functions for $\delta / r < 0.1$ leads to a very good estimate for values of $\delta / r$ until 1.

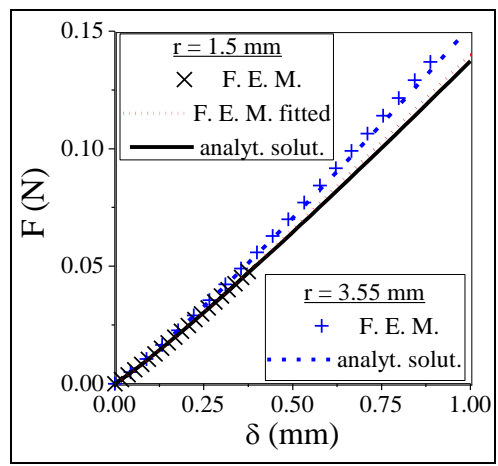

**Figure 4.** *Comparison of the force vs. indentation depth curves for FEM and analytical solution given in Appendix 1 corresponding to a membrane at a constant tension $t = 107\ N / m$ and $D = 102\ mm$.*

### *2.3. Contact condition (Hertz Problem)*

The contact Hertz problem of a rigid sphere indenting an elastic semi-infinite medium leads to the following analytical prediction for a frictionless contact:

$$F_H = (4/3) \ E^* \ r^{0.5} \ \delta^{1.5} \qquad (1)$$

where $E^* = E / (1 - \nu^2)$, E and $\nu$ are respectively the Young's modulus and Poisson's ratio of the medium, r the radius of the indentor, $\delta$ the indentation depth and $F_H$ the reaction force acting on the indentor.

The two extreme cases, i. e. 0° and 90°, are examined for the friction angle between the indentor and the solid surface. Considering the force versus indentation depth, the difference between the frictionless case and adhesive case is not significant whatever the ratio between surface tension and elastic modulus is. This insensitivity to the contact condition is well known for the Hertz contact problem[11]. The Figure 5 shows the ability of the FEM to confirm this result, at least for low indentation depths ($\delta$ / r < 0.25). The difference between the FEM and the analytical result at low contact areas ($\delta$ / r < 0.05) can easily be reduced by refining the mesh. The assumption of the insensitivity to the contact condition is still valid when accounting for surface tension in the Hertz contact problem or even considering the "drum" problem of Appendix 1. Indeed, Figure 6 corresponds to a the force vs. indentation depth curve for the following parameters: t = 0.03 N/m, E = 1 MPa, $\nu$ = 0.42, r = 50 nm, D = 100 r and h = 30 r. The mechanical response is dominated by the surface tension since the reaction force at $\delta$/r = 0.125 is approximately 3 times the prediction of Hertz ($F_{\delta/r=0.125} \approx 3 \ F_{H, \ \delta/r=0.125}$). The difference between the computation considering a frictionless contact and that considering an adhesive contact is always less than $10^{-3}$. Figure 6. also shows that the force vs. indentation depth curve is well fitted by a power law function $F \approx 3.7 \ \delta^{1.2}$ for this particular case.

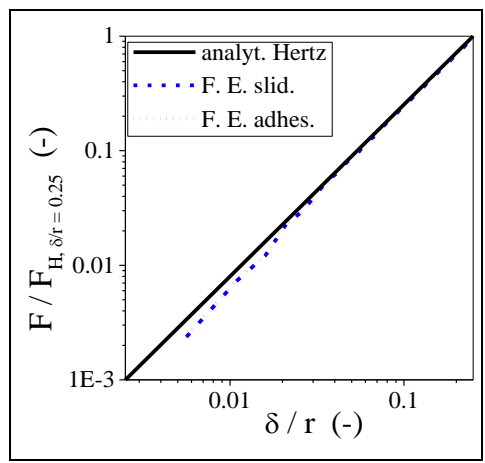

**Figure 5.** *Comparison of the force vs. indentation depth obtained with the FEM for perfectly sliding contact and a perfectly adhesive contact with the analytical solution of the Hertz contact problem.*

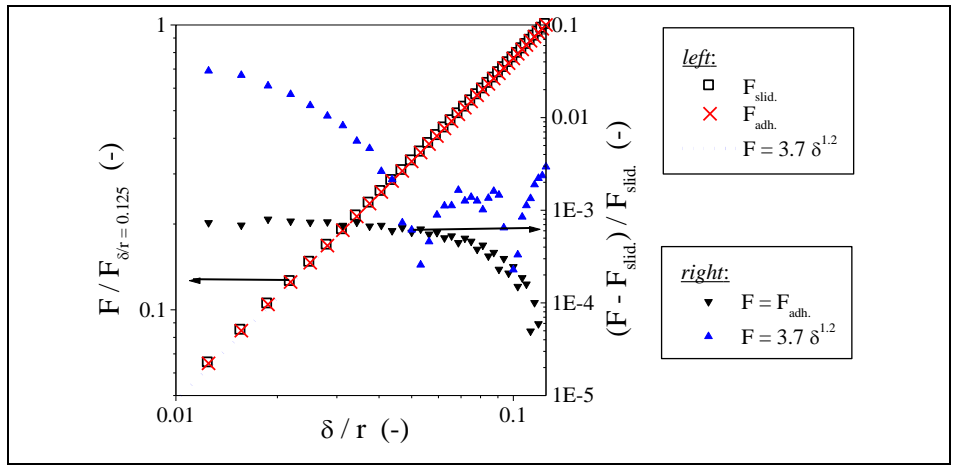

**Figure 6.** *Comparison of the force vs. indentation depth obtained with the FEM considering a frictionless contact (fitted by a power law curve) and a perfectly adhesive contact corresponding to a mechanical response dominated by the surface tension.*

### 2.4. Border effect

The analytical solution derived in Appendix 1 indicates that, for a given indentation depth, the reaction force acting on a stretched membrane tends to zero when the diameter D tends to infinity. The relative sizes D / r and h / r of the FEM must be sufficiently large to account for this tendency, especially at large values of surface stiffness modulus, i. e. typically at t / r E > 1. The Figure 7 quantifies the border effect as a function of the size of the FEM For high surface stiffness moduli, the relative size D / r of the FEM should be at least 300 in order to reach the asymptotic value and h must be of the order of magnitude of D.

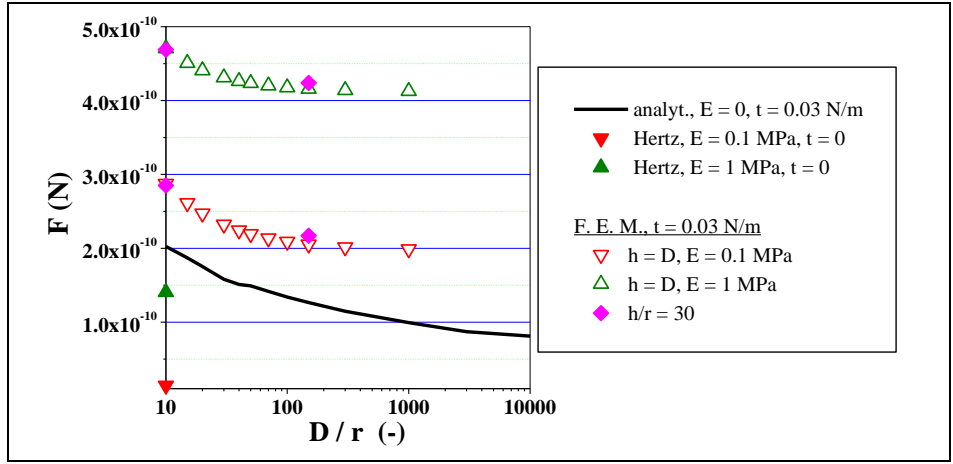

***Figure 7.*** *Force vs. size of the FEM for r = 50 nm and δ = 5 nm.*

## 3. Assessment from experiments

### 3.1. Indentation of a stretched rubber membrane

In order to validate the numerical computations, an indentation test using a pre-stretched rubber membrane and steel balls, shown in Photo 1, has been performed. The indenter is a sphere and the membrane is initially a planar disc. For large indentation depths, the tension in the membrane increases slightly due to the stretching of the rubber. Nevertheless, since the pre-stretch ratio is 4 in bi-axial tension, it is expected that the tension in the membrane is almost constant during the test. The rubber is an industrial grade of rubber for balloons. The tension in the membrane is estimated by measuring the curvature of the pressurized membrane corresponding to the gravity acting on 10 cm of water. The accuracy of this measure is ±5 %. Both dry and lubricated contact conditions are shown in Figure 8, for two radii of the indenting sphere. Finite elements results are compared to experimental results. The power law curves

$$F \approx a \, \delta^b \qquad (2)$$

where δ is the indentation depth, a and b depend on material constants and indentor radius and F is the force acting on the indentor. The constants a and b are obtained by fitting the numerical results at small indentation depths, typically δ < r / 4. Since the tension in the membrane is known in a range ±5 %, the error bars have been associated to the numerical predictions. Numerical results show a very good agreement with experiments in a relatively large range of the δ / r ratio. Experiments confirm that the contact conditions, frictionless or not, can be seen as a second order parameter. Since less numerical convergence problems appear for frictionless contact and owing to the insensitivity to the contact condition, all the following computations have been performed with this assumption.

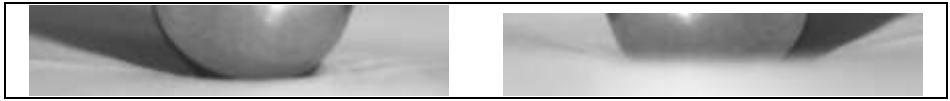

**Photo 1.** *Photo of a metallic sphere indenting a rubber membrane.*

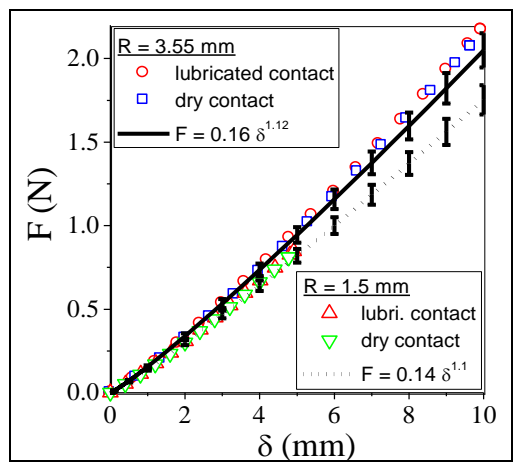

**Figure 8.** *Force vs. indentation depth corresponding to Photo 1.* *t = 420 N / m ±5 %, D = 102 mm.*

### 3.2. AFM indentation of soft materials

The material used is a polydimethylsiloxane (PDMS). No substrate effects are expected when nano-indenting the polymer since the thickness of the PDMS substrate is bigger than 1 mm. In this study, only the surface exposed at the air during crosslinking is considered. This means that no preferential crosslinking at the surface is expected when compared with crosslinking in the bulk[5].

One of the fundamental points to obtain reproducible, quantitative and reliable data is the calibration procedure, which should be rigorous and systematic for all measurements. This calibration procedure is fully described in the literature. In particular, the spring constant of the cantilever has been determined by using a nondestructive method, based on the use of reference rectangular cantilevers. The cantilever used in this study was a triangular shaped cantilever (supplied by Nanosensor- Germany) and had an effective $0.30\pm 0.03$ N/m This value is similar to the one determined by the thermal fluctuation measurement method.

Moreover, the determination of the Young's modulus by a nano-indentation experiment assumes the knowledge of the contact area value between the indentor and the sample. The Figure 9 shows a scanning electron microscope (SEM) image of an atomic force microscopy tip (AFM) from which the radius is estimated to be about 46 nm in the image plane and assumed to be spherical since it is difficult to know the exact shape of such tips.

The AFM force-indentation depth curves are deduced from the AFM force curves obtained in the quasi-static mode (the tip does not oscillate). In particular, no creeping effect and no plastic behaviour have been observed during the indentation experiments. This means that the PDMS used herein behaves like perfectly elastic material during the nano-indentation experiment (in considering that the indentation depth is about 25 nm). Moreover, the beginning of the indentation is assumed to correspond to the minimum of the force in the force vs. piezo displacement curve. Finally, dynamic mechanical tests have shown that the Young's modulus, about 0.13 MPa, is quite independent of the deformation rate and is in the range of the rubbery plateau within the experimental conditions.

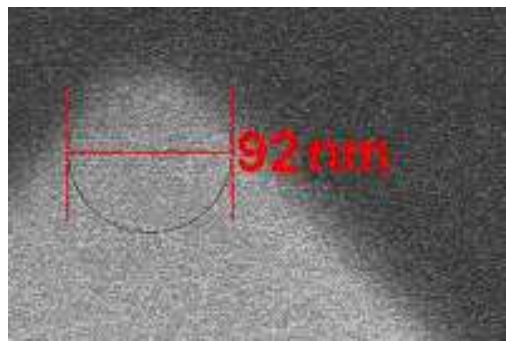

**Figure 9.** *SEM image showing the projection in a plane of an AFM tip (experimental data's from O. Noël[5]).*

## 4. Young's modulus determination using the present model

The Figure 10 shows force vs. indentation depth curves corresponding to a nano-indentation experiment of the material described in the above section using the AFM tip shown in Figure 9, compared to the Hertz's prediction and to the present model. The model assumes a spherical AFM tip which radius is approximately equal to 46 nm. The bulk modulus is assumed to be 2 GPa. It is expected that the surface tension of the material is in the range 0.02 N/m to 0.025 N/m. Three FEM computations are compared corresponding to t = 0, i.e. Hertz's prediction, t = 0.0225 N/m and t = 0.03 N/m. It appears clearly that it is necessary to account for surface tension to retrieve the experimental data's and obtain a reaction force in the expected range.

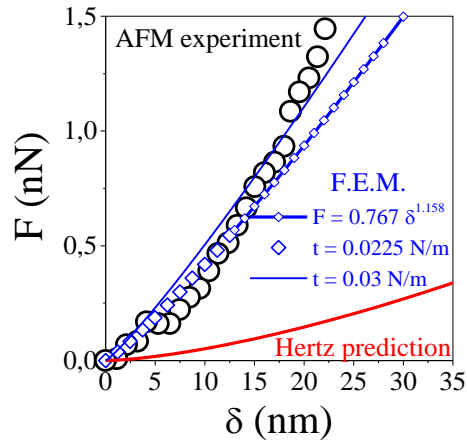

**Figure 10.** *Force vs. indentation depth corresponding to AFM tip shown in Figure 9 (experimental data's from O. Noël[5]).*

## 5. Summary

Whenever the present model is valid, it is then possible to estimate the surface tension from a fit of the experimental force vs. indentation. Indeed, the exponent b is directly related to the surface stiffness modulus t / (r E*). For the sake of simplicity, the Figure 11 proposes an estimate of the b exponent in the case of quasi-incompressible materials, using a Boltzman function:

$$\frac{t}{r\,E^*} \approx 0.2 \ \{\frac{0.41}{b-1.09} - 1\}^{1.15} \tag{3}$$

For low surface stiffness moduli, typically t / (r . E*) < 0.01, the Hertz's prediction prevails, i. e. b ≈ 1.5. At larger values of the surface stiffness moduli, it is necessary to account for surface tension, at least for low indentation depths since the present results correspond to δ / r < 0.3.

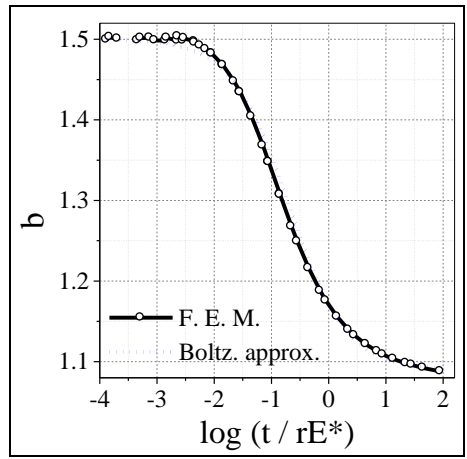

**Figure 11.** *b exponent vs. surface stiffness modulus, Boltzman Fit and numerical results.*

## 6. Discussion

The reaction force acting on an indenter depends on its shape and size. Considering AFM nano-indentation, it is relatively difficult to measure and well define the three dimensional geometry of the AFM tip and also to be sure that the tip is clean. ~~Cleanliness of the AFM tip has been regularly verified in doing force curve measurements on a reference wafer of silicon. Consequently, the tip is expected to be contaminated if the measured force after a nano-indentation experiment is different from the one measured on the reference silicon wafer before the nano indentation experiments. However,~~ the difference between the assumed shape and size and the real ones could be responsible for the significant discrepancies between several simulations and experiments. In order to verify this point, macroscopic indentation experiments are under study.

Moreover other surface forces exists. Assuming that, for instance, only surface tension is acting render the material properties dependant on the mechanical model. The interaction between the AFM tip and the substrate, such like adhesion, can also influence, in addition to the surface tension of the substrate, the force vs. indentation depth response. However, it is relatively simple to account for these kind of surface forces in the FEM model once those forces are known in magnitude and direction, the major technical difficulty being to account for surface tension.

Finally, the Figure 12 shows that a shift of time assumed to correspond to the beginning of the indentation influences significantly the slope of the force vs.

indentation depth curve. Hence, the deduction of the material properties from such experiments is not easy if this time is not well defined.

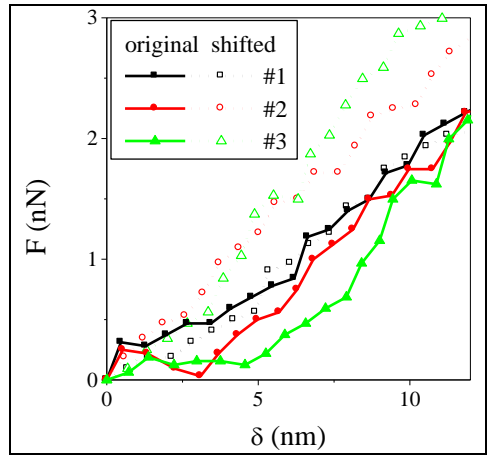

**Figure 12.** *Influence of the time corresponding to beginning of the indentation on three force vs. indentation depth curves (experimental data's from O. Noël[5]).*

The Figure 13 shows typical force vs. indentation curves for several AFM tip radii and Young's moduli in the range concerned with surface tension. Table 1 presents the a and b material constants of eq. 2 corresponding to the fitting of the curves of Figure 13 in the range $0 < \delta < r / 4$. For these experiments, the triplet (AFM radius tip, Young's modulus and surface tension) is known. Nevertheless, the deduction of the third parameter assuming the knowledge of two of them lead to a possible use of the model. The model accounting for elasticity and surface tension gives results in the right range. But the discrepancy that clearly appears indicating that it should be completed. The effect of adhesion should be quantified and the precision on the time corresponding to the beginning of indentation increased.

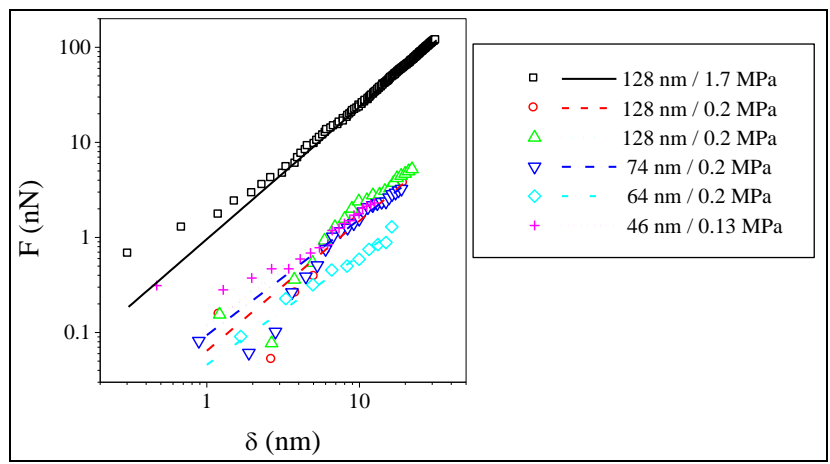

**Figure 13.** *Typical force vs. indentation curves for several AFM tip radius and Young's moduli.*

| Material characteristics | | | Power law fit | | Deduced from AFM indentation | | |
|---|---|---|---|---|---|---|---|
| r (nm) | E (MPa) | t (N / m) | a (N.m$^{-b}$) | b | r$_{\mid E,t}$ (nm) | E$_{\mid t,r}$ (MPa) | t$_{\mid E,r}$ (N / m) |
| 128 | 1.7 | 0.022 | 3.05 10$^{+3}$ | 1.39 | 155 | 3.7 | 0.019 |
| 128 | 0.2 | 0.022 | 127 | 1.37 | 993 | 2.7 | 0.003 |
| 128 | 0.2 | 0.022 | 11.7 | 1.22 | 180 | 0.49 | 0.016 |
| 74 | 0.2 | 0.022 | 8.5 | 1.22 | 173 | 0.81 | 0.010 |
| 64 | 0.2 | 0.022 | 0.98 | 1.15 | 53 | 0.29 | 0.027 |
| 46 | 0.13 | 0.022 | 6.0 | 1.19 | 177 | 0.89 | 0.006 |

**Table 1.** *Deduction of the third parameter assuming the knowledge of two of the triplet (AFM radius tip, Young's modulus and surface tension).*

## 7. Conclusion

Although the accuracy of the numerical results of the FEM proposed herein is expected to decrease with the rotation angle of the surface elements, it is in good agreement with the precision available for experimental data at nanoscales in most cases. This FEM allows then to account relatively simply for surface tension in contact problems by using a standard finite elements free software and does not require the implementation of specific procedures. Nano-indentation can be a useful tool to measure the surface tension of rubbers or bio-material once a convenient mechanical model allows to consider the surface forces acting at this scale.

## Appendix

### A1. The "drum" problem

The analytical solution for the indentation of a stretched membrane is derived below. The biaxial tension of the membrane is denoted t and assumed to be independent of the stretch induced by the indentation so that the problem remains purely geometrical. This practically corresponds to membranes so highly stretched that the small strains induced by the indentation does not change significantly the stress state. The indenter is assumed to be spherical, as shown in Figure A1. 1, the membrane is initially plane and gravity is not considered. The quasi-static equilibrium leads to:

$$F = 2 \pi t \sin \alpha(D) \qquad (a1.\ 1)$$

where F is the applied force on the spherical indenter, D the distance from the axis of symmetry and $\alpha(D)$ the angle between the deformed and initial state at a distance D.

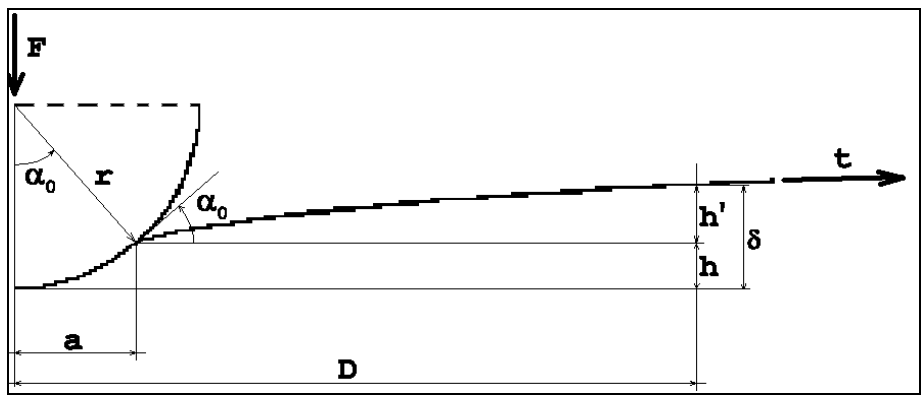

*Figure A1. Indentation of a membrane with a sphere.*

The radius of the spherical indenter and the contact area are denoted respectively r and a, as shown in Figure A1. The membrane is tangent to the sphere at the contact point so that:

$$F = 2 \pi t \, a^2/r \qquad (a1.\,2a)$$

$$\delta = h + h' = r - \sqrt{r^2 - a^2} + \int_a^D \frac{a^2/r\rho}{\sqrt{1 - (a^2/r\rho)^2}} \, d\rho \qquad (a1.\,2b)$$

where $\delta$ denotes the indentation depth. The last integral leads to a hypergeometric function which tends to infinity when D tends to infinity. The relation between the applied force and the indentation depth cannot be simply derived from the last equations but an approximation for a << r is given by:

$$\delta \approx \frac{F}{6\,t} \left\{ \ln(D) - [\ln(r) - \ln(\frac{F}{2\,\pi\,t})]/2 \right\} \qquad (a1.\,3)$$

The Figure A1. 2. shows the comparison between the exact solution and this approximation. In all cases, the error is less than 12 % for $\delta / r < 1$. This Figure also shows that, for a given normalized indentation depth, the contact area is dependant on the size of the membrane.

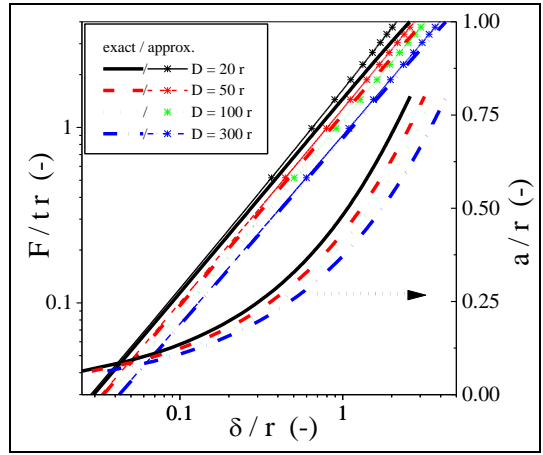

**Figure A2.** *Comparison of the exact solution given by eq. (a1. 2) approximation of eq. (a1. 3).*